\documentclass[12pt]{article} 
\usepackage{psfig}  \usepackage{epsfig}  \usepackage{graphics}
\usepackage{glas}  
\def\Journal#1#2#3#4{{#1} {#2} (#4) #3 }

\def\NPB{{ Nucl. Phys.} B}
\def\PLB{{ Phys. Lett.}  B}

\def\ZPC{{ Z. Phys.} C}
\def\lsim{\mathrel{\rlap{\lower4pt\hbox{\hskip1pt$\sim$}}
    \raise1pt\hbox{$<$}}}                
\def\gsim{\mathrel{\rlap{\lower4pt\hbox{\hskip1pt$\sim$}}
    \raise1pt\hbox{$>$}}}                
\newcommand{\Herwig}{{HERWIG}}

\newcommand{\GeV}{\mbox{\rm ~GeV}}

\begin{document}
\begin{titlepage}{GLAS-PPE/1999--12}{September 1999}
\vspace*{10mm}
\title{
A Comparison of Deep Inelastic Scattering \\
Monte Carlo Event Generators 
 to HERA Data}
\author{
N.~H.~Brook\Instref{glas}, T.~Carli\Instref{MPI}, E.~Rodrigues\Instref{Bristol}, M.~R.~Sutton\Instref{oxford}, N.~Tobien\Instref{DESY}, M.~Weber\Instref{heid}}
\Instfoot{glas}{ Dept. of Physics \& Astronomy, University of Glasgow, Glasgow, UK.}
\Instfoot{MPI}{Max-Planck-Institut f\"ur Physik, M\"unchen, Germany}
\Instfoot{Bristol}{H.H.~Wills Physics Laboratory, University of Bristol, Bristol, UK.}
\Instfoot{oxford}{Department of Physics, University of Oxford, Oxford, UK.}
\Instfoot{DESY}{ DESY, Notkestra\ss e 85, Hamburg, Germany.}
\Instfoot{heid}{Institut f\"ur Hochenergiephysik, Universt\"at Heidelberg, Heidelberg, Germany.}
\begin{abstract}
 The Monte Carlo models ARIADNE, HERWIG and LEPTO
 are compared to deep-inelastic scattering data
 measured at the ep-collider HERA.
\end{abstract}
\vfill
\conference{contribution to the \\
 1998-1999 HERA Monte Carlo workshop.}
\end{titlepage}

\section{Introduction}

Monte Carlo generators are an essential tool in modern day experimental
High Energy Physics. They play a crucial r\^{o}le in the analysis of the
data, often in assessing the systematic errors of a measurement. 
For that reason
it is of great importance that the Monte Carlo programs give results
that agree closely with the experimental data.
This paper aims to describe the agreement, deficiencies and tuning of
the Monte Carlo models with the neutral current
deep inelastic scattering (DIS) data at HERA. 
Extensive use is made of the
utility package HzTool~\cite{hztool}, which is a FORTRAN library containing a
collection of experimental results from the H1 and ZEUS collaborations.

The work described here is part of an ongoing program. During
the workshop a forum was established between the H1 and ZEUS
collaborations for a joint coordinated investigation of the generators
working closely with the programs' authors.

\section{Monte Carlo Models}
The ARIADNE~\cite{ariadne}, HERWIG~\cite{herwig} and LEPTO~\cite{lepto}
 Monte Carlo generators for DIS data have
been investigated during the course of the workshop. 
Other programs such as RAPGAP~\cite{rapgap} and
those developed over the duration of the workshop will be examined as
part of the ongoing program of work.
In the following sections a brief introduction to each
of the three generators studied is given.

\subsection{ARIADNE}

 In ARIADNE the QCD cascade is
 modelled by emitting gluons from a chain of independently radiating
 dipoles spanning colour connected partons~\cite{dipole}, correcting the
 first emission to reproduce the first order matrix elements~\cite{MEref}.
 The hadronisation of the partons into final state particles is 
performed by
 the Lund string model~\cite{string}
 as incorporated in JETSET~\cite{jetset}.
 Since the proton remnant
 at one endpoint of the parton chain
 is treated as an extended object, the coherence condition
 allows only a fraction of this source
 to be involved in gluon radiation.
 Since the photon probing the proton only resolves the struck
 quark to a distance $\lambda \sim 1/Q,$
 the struck quark is also treated as an extended object.
 As a consequence gluon emissions in the proton
 and photon directions are suppressed.
 This phase space restriction is governed by
 $a={(\mu/k_T)}^\alpha$ where $k_T$ is the transverse momentum of the
emission, $a$ is the fraction of the
 colour antenna involved in the radiation,
 $\mu$ is a parameter related to a typical inverse size
 of a hadron and $\alpha$ governs the distribution of
 the energy along the dipole.
 
In the default version of 
ARIADNE, the mechanism for soft suppression of radiation
due to the extended source of the proton remnant results in a
suppression of radiation in the current region of the
Breit frame at high $Q^2.$ In the course of the workshop
a high $Q^2$ modification was developed~\cite{hiq2mods}
 where this suppression in the current region was removed.

\subsection{HERWIG}

HERWIG relies on a coherent parton branching algorithm
with additional first order matrix element corrections \cite{seymour}
to populate the extremities of phase space which the partons from
the conventional QCD cascade fail to occupy.
The partons are transformed into hadrons using the cluster
fragmentation model \cite{cluster}, whereby the primary hadrons are
produced from
an isotropic two body decay of colour-singlet clusters formed from
partonic constituents.

Since the Monte Carlo tuning to HERA data at the
`Future Physics at HERA workshop'~\cite{futurephys}, a new version  of 
HERWIG (version 5.9) has become available.
This version includes the  modified remnant treatment of version 5.8d
whereby the fragmentation of the cluster containing the hadronic remnant
is treated
differently to  that containing the perturbative parton from the
incident hadron. In addition, the particle decay tables have been
updated and now contain a large amount of information on additional resonance
decays.

The default version of \Herwig\ implements the next-to-leading order (NLO)
running of the QCD coupling constant $\alpha_s$. The \Herwig\ philosophy
is to incorporate as much perturbative QCD behaviour as possible, so even 
though the generator only uses a leading order (LO)
parton shower cascade,  a NLO $\alpha_s$ behaviour is implemented. 
This can be justified, to some degree, because of the HERWIG
implementation of angular ordering in the QCD cascades.
The H1 collaboration have modified \Herwig\ to allow a LO behaviour of
$\alpha_s$ \cite{chyla}.

\subsection{LEPTO}

In LEPTO  the hard parton processes are described by a
leading order matrix element (ME). The soft and collinear divergences
are regulated with a lower and upper cut in $z_{p}$ where
$z_p= p \cdot j_1 / p \cdot q$ where $p$ (q) is the proton (photon)
four-vector and $j_1$ the four-vector of one of the partons
produced in the hard subprocess. In addition, the
invariant mass squared of the two hard partons
is required to exceed a minimal value, $\hat{s}_{min}$.
Below the ME cut-offs,
parton emissions are treated by parton showers
based on the
DGLAP evolution equations \cite{DGLAP}.  The amount of parton radiation
depends on the virtuality chosen between a lower cut-off
($Q_0^2$) and a maximum given by the scale of the hard
process or the ME cut-off.
LEPTO uses JETSET for the hadronization of the partons. In addition to
this non-perturbative phase, LEPTO introduces another non-perturbative
mechanism.
This is a soft (i.e.~at a scale below $Q_0^2$) colour
interaction which assumes that the colour configuration of the
partonic system can be changed whilst traversing
 the colour field of the proton remnant.
This was introduced in order to reproduce the rapidity gap events
observed at HERA.

During the course of the workshop a new version of LEPTO was released
(version 6.5.2$\beta$). This version introduced a new scheme for dealing
with SCI events, in which the
probability of soft colour interactions is suppressed depending on the
difference in area spanned by the possible string configurations (after
or before a soft colour interaction)~\cite{rathsman}.
This means at high $Q^2$ there are effectively no
soft colour interactions.

\section{Model comparison with the data}

\subsection{ARIADNE}

These studies closely followed those of the previous tuning exercise
\cite{futurephys} performed at the `Future Physics at HERA' workshop. 
However, they have been extended: jet
data were now available for inclusion in the comparisons; and, the
behaviour of the parameter PARA(25) was considered for the first time.
ARIADNE version 4.10 has been investigated including the modified treatment
of high $Q^2$ DIS events (MHAR(151) = 2.)

\begin{table}[htbp]
\centering
  \begin{tabular}{|c|c|c|c|l|}
  \hline
  Parameter & Default & \multicolumn{2}{|c|}{Tuned} & Description \\
  & & set 1 & set 2 & \\
  \hline
  \hline
  PARA(10) & 1.0 & 1.6 & 1.2 & power in soft suppression for the remnant \\
 \hline
  PARA(15) & 1.0 & 0.5 & 1.0 & power in soft suppression for the struck quark
\\
  \hline
  PARA(25) & 2.0 & 1.4 & 1.5 & governs probability of emissions outside\\
           &     &     &  &the soft suppression cut-off\\
  \hline
  PARA(27) & 0.6 & 0.6 & 0.6 & square root of primordial $k^{2}_{T}$ in
GeV\\
  \hline
  \end{tabular}
\caption{{\it Values of the ARIADNE parameters before and after tuning.}}
\label{tab:parameters}
\end{table}

The four model parameters considered are listed in Table
\ref{tab:parameters}, which includes a short description of their
influence.
The two parameters PARA(10) and PARA(15) govern the slope of the
suppression
line (in the phase space available for gluon emission) for the proton
and the struck quark, respectively. PARA(25) governs the probability of
emissions
outside the soft suppression cut-off, while PARA(27) corresponds to the
square root of the primordial $k^{2}_{T}$ in the proton.

\subsubsection{Approach 1}
This approach concentrated on HERA data at $Q^2 > 80{\rm\ GeV^2}.$
The motivation for this was to
minimize the theoretical uncertainties in the generator associated with
parton evolution in the low $(x,Q^2)$ region.
The distributions that were most sensitive to the parameters under
investigation were first ascertained. 
Next a combined $\chi^2$ was calculated for
each parameter setting according to
\begin{equation}
  \chi^{2}_{Comb}=\frac{1}{nsets}\sum_{i=1}^{nsets}{\chi^{2}_{i}}  
\label{eq:chi_sq_comb}
\end{equation}
where $\chi^{2}_{i}$ represents the total (average) $\chi^{2}$ per
degree of freedom (d.o.f) of data set $i$.
The parameter combination that yields the minimum of the overall
$\chi^{2}_{Comb}$ corresponds to the tuned result.

The following distributions for $Q^2 > 80\ {\rm GeV^2}$ were investigated:

\begin{itemize}
  \item   scaled momentum $x_p$ distributions in the current region
of the Breit frame \cite{jane};
  \item  flows of transverse energy in the hadronic centre of mass
              system \cite{hiq2e};
  \item  differential distributions and evolution of mean of event
              shape variables thrust $T_c$ and $T_z$, jet broadening $B_c$
               and jet mass $\rho_c$ in the current region of the Breit
              frame \cite{97098};
  \item  fragmentation functions and charged particle multiplicities
               in the current region of the Breit frame \cite{97108};
  \item  differential and integrated jet shapes as a function of
               pseudorapidity $\eta$ and transverse energy $E_T$ \cite{98038};
  \item  (2+1) jet event rate as a function of the transferred
               momentum squared,$Q^2$ \cite{98087}.
\end{itemize}

\begin{table}[htbp]
\centering
  \begin{tabular}{|c|c|c|}
  \hline
  Data set ref. & $\chi^2_{default}$ & $\chi^2_{tuned}$ \\
  \hline
  \hline
  \cite{jane}  &  1.94 &  2.60 \\
  \hline
  \cite{hiq2e} &  0.85 &  0.65 \\
  \hline
  \cite{97098} &  1.69 &  1.70 \\
  \hline
  \cite{97108} &  0.58 &  0.65 \\
  \hline
  \cite{98038} & 17.34 & 12.49 \\
  \hline
  \cite{98087} &  8.10 &  3.08 \\
  \hline
  \hline
  $\chi^{2}_{Comb}=\frac{1}{nsets}\sum_{i=1}^{nsets}\chi^{2}_{i}$
        &  5.08 &  3.53 \\
  \hline
  \end{tabular}
\caption[junk]{{\it Summary of the $\chi^2$ values obtained before and
after tuning for all sets of hadronic final state data.}}
\label{tab:summary}
\end{table}

Table \ref{tab:summary} summarises the total
$\chi^2$ for each of the six sets of data, together with the combined
$\chi^{2}_{Comb}$ given by equation \ref{eq:chi_sq_comb}.
The results of tuning the parameters, set 1 Table~\ref{tab:parameters},
agree very well with those previously
obtained \cite{futurephys}. Both transverse energy flows
and the jet data 
strongly favour high values for PARA(10), in contrast to the lower value
favoured by the charged particle momentum distributions .
Again the value PARA(15)=0.5 yields a better overall description of
data, as
compared to its default value of 1.0. The jet variables are particularly
sensitive to PARA(15). The behaviour of PARA(25) was studied for the
first
time. Although its influence is not large in general, it clearly has a
significant effect on the predictions related to the (2+1) jet event
rate and on the transverse energy
flows. The results suggest a lower value compared to the default one.
Parameter PARA(27) is a relatively insensitive parameter, but the data
disfavour high values, such as 0.8-1.0 GeV, when describing transverse
energy flows and jet shapes.
The tuning has been performed using the GRV94 parton density function
\cite{grv94}. The use of the parton density function CTEQ4M~\cite{cteq4}
results in a slightly worse value for $\chi^{2}_{Comb}$.

\begin{figure}[htbp]
\centering
\epsfig{file=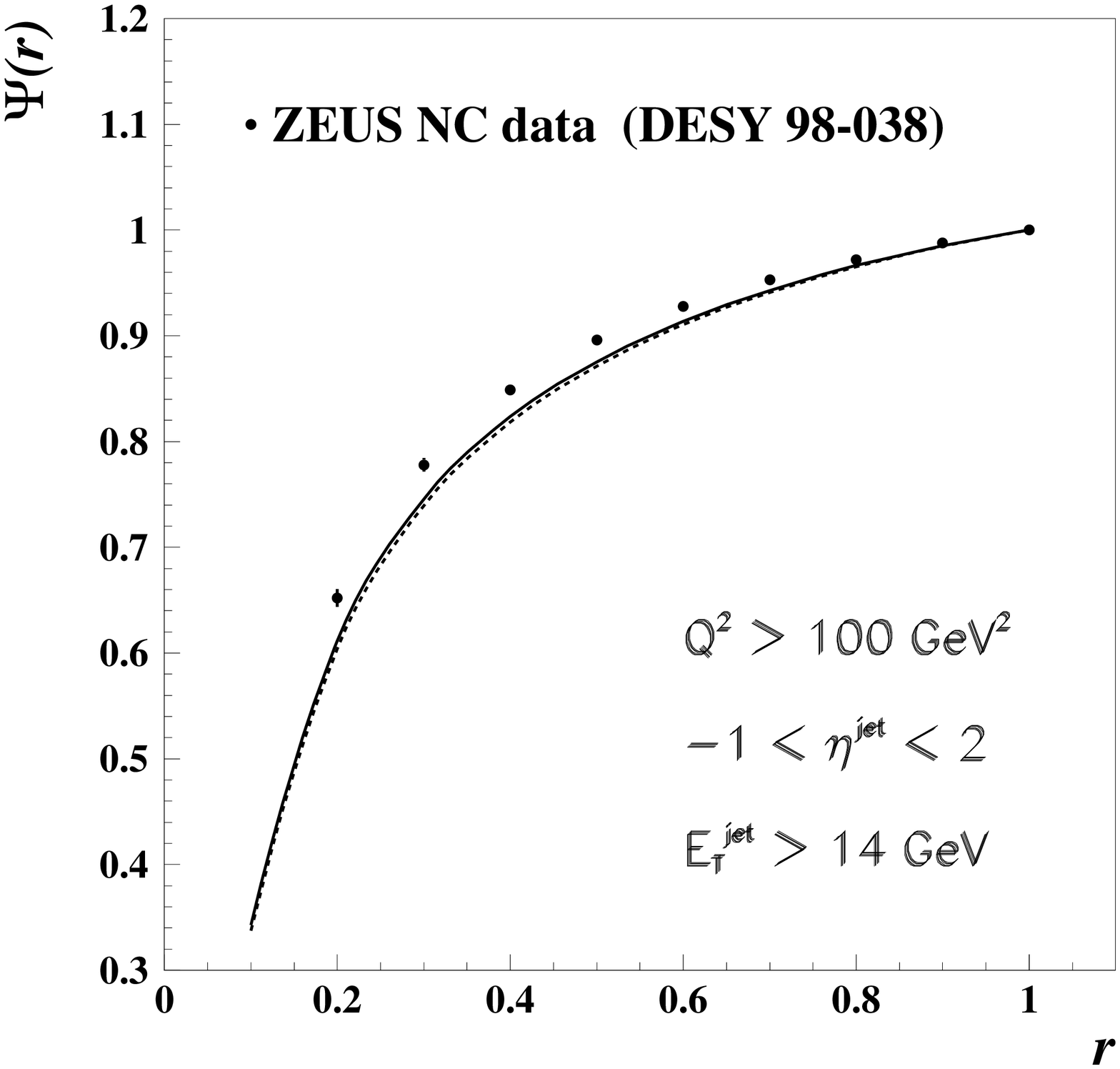,height=0.4\textheight}
\caption[junk]{{\it The integrated jet shape $\Psi(r)$ before (dashed
line) and after (solid line) tuning ARIADNE (approach 1).}}
 \label{fig:98038}
\end{figure}

\begin{figure}[hbtp]
\centering
\epsfig{file=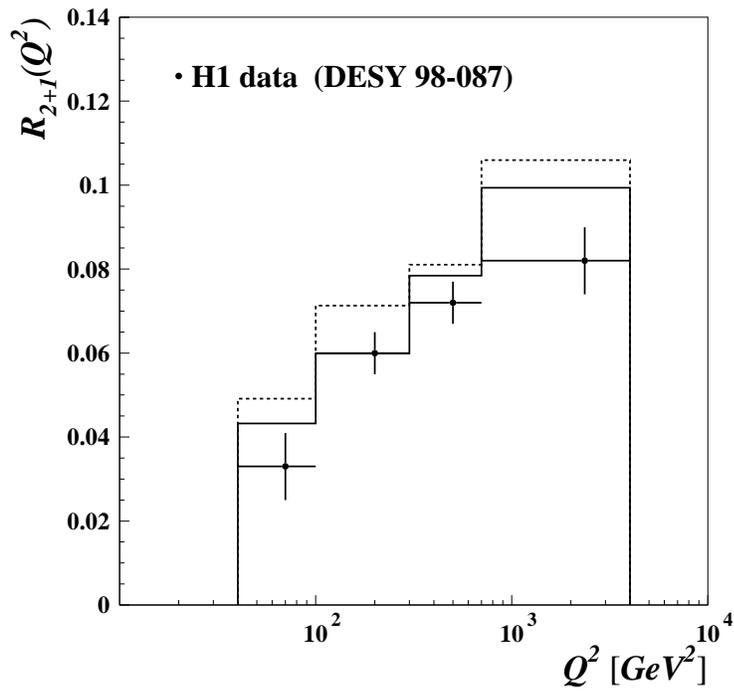,height=0.4\textheight}
\caption[junk]{{\it The (2+1) jet event rate $R_{2+1}(Q^2)$ before
(dashed line) and after (solid line) tuning ARIADNE (approach 1).}}
 \label{fig:98087}
\end{figure}

\begin{figure}[htbp]
\centering
\epsfig{file=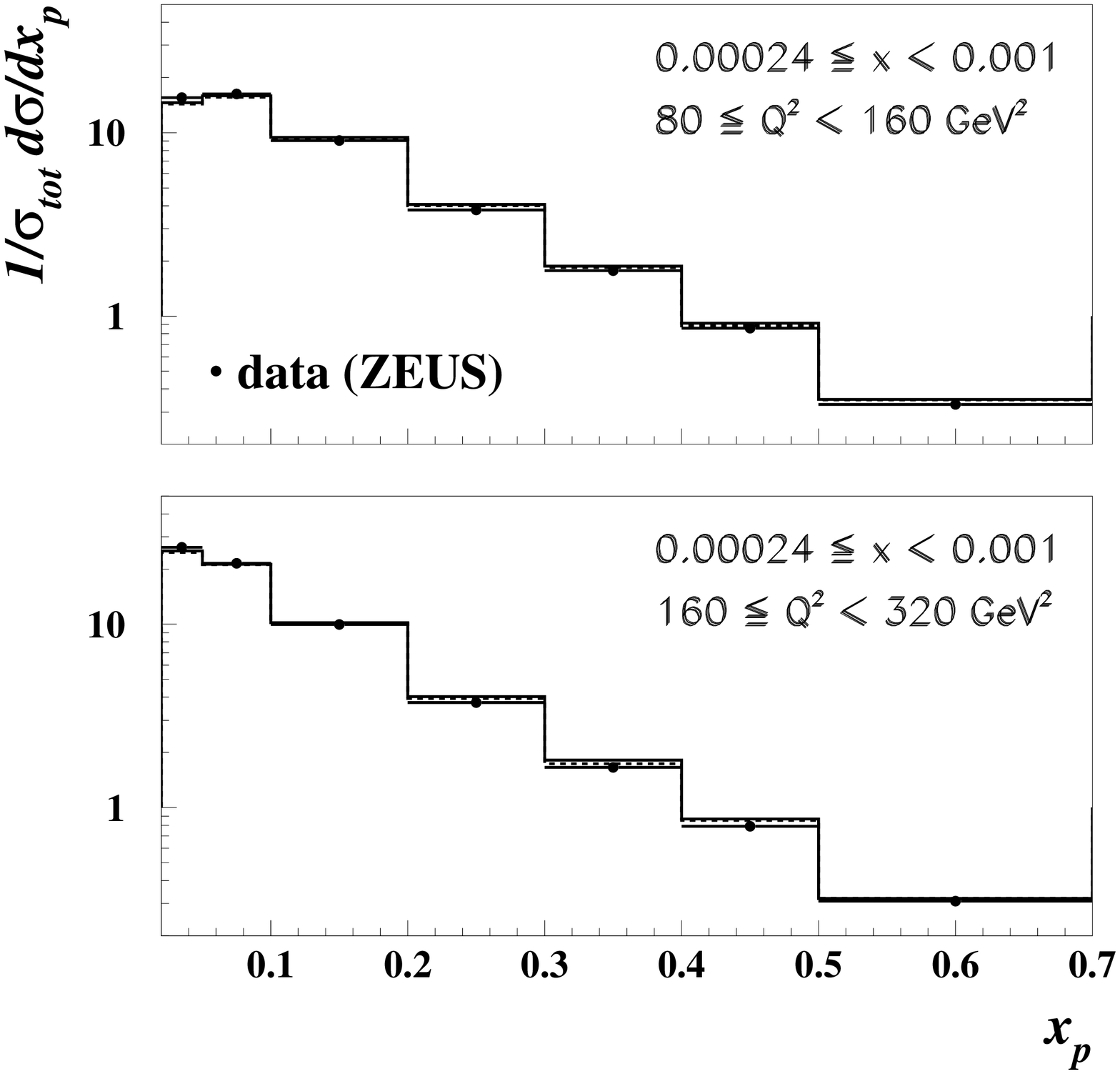,height=0.45\textheight}
\epsfig{file=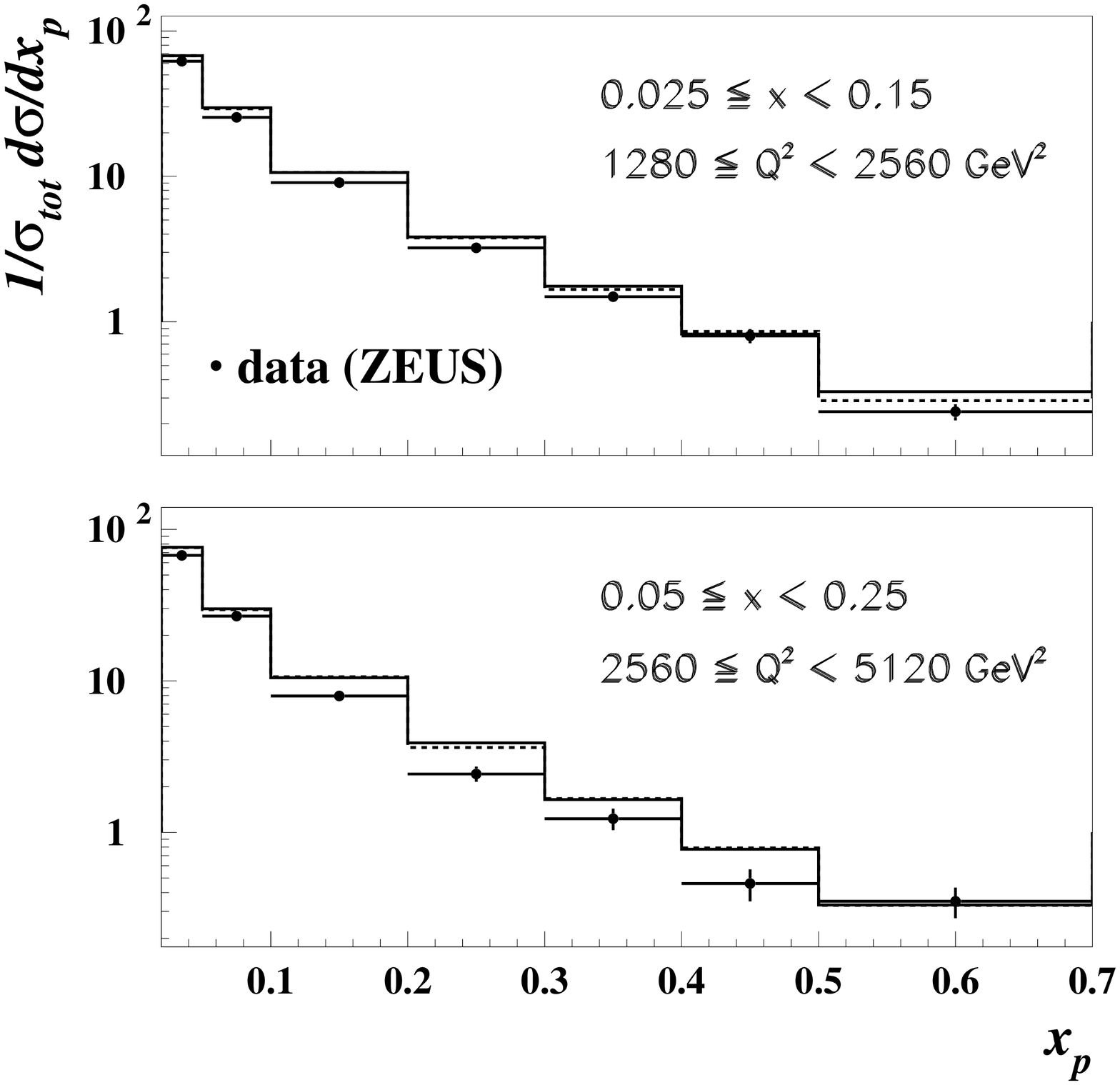,height=0.45\textheight}
\caption[junk]{{\it The scaled momentum $x_p$ distributions in the current
region of the Breit frame before (dashed line) and after
(solid line) tuning ARIADNE (approach 1).} }
 \label{fig:jane}
\end{figure}

\begin{figure}[htbp]
\centering
\epsfig{file=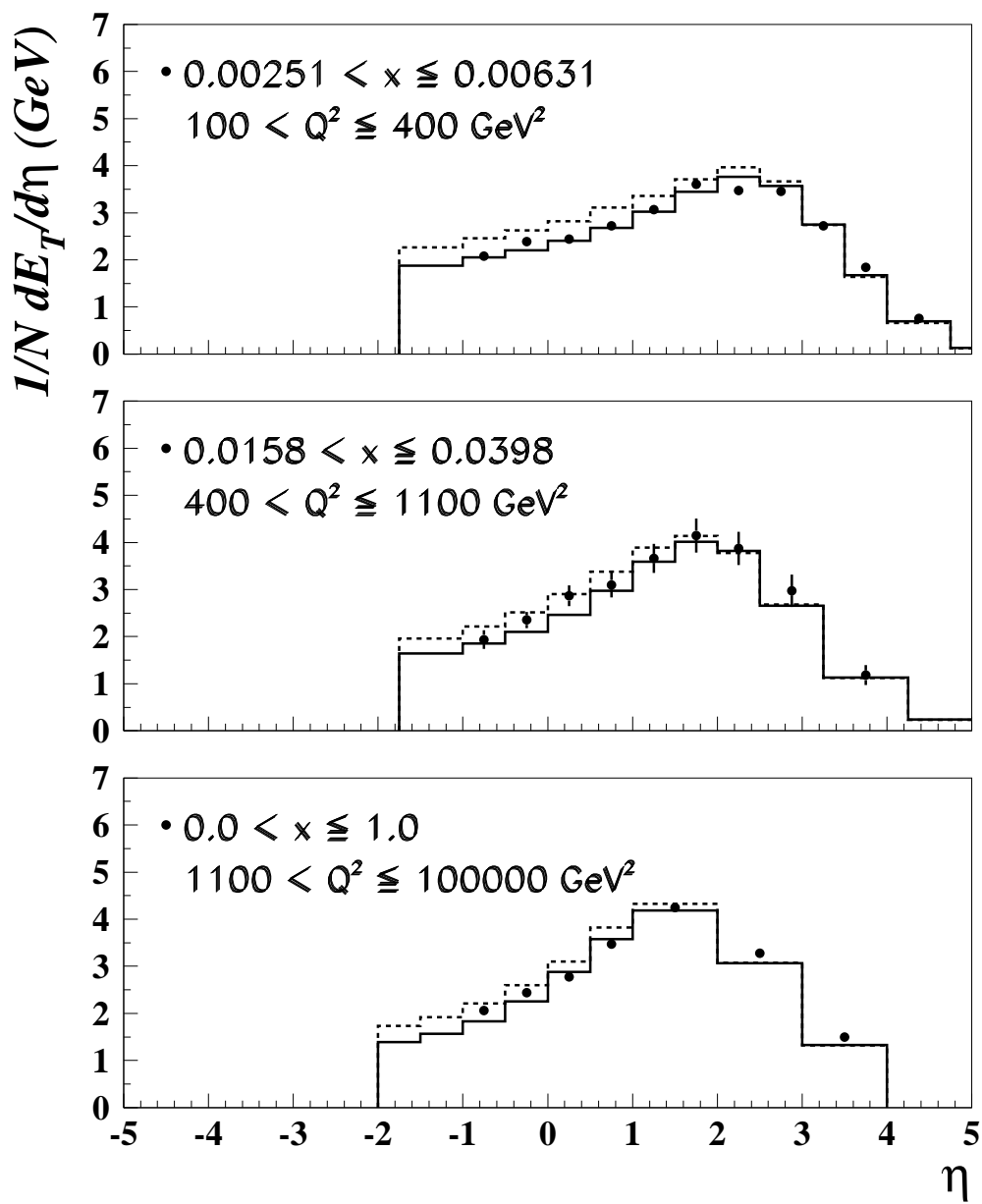,width=0.9\textwidth}
\caption[junk]{{\it H1 preliminary transverse energy flows in the hadronic
center-of-mass system before (dashed line) and after (solid line)
tuning ARIADNE (approach 1). }}
 \label{fig:hiq2e}
\end{figure}

The improvement achieved with the tuned parameters is mostly due to a
better
description of jet shapes and (2+1) jet event rates (see Figs.
\ref{fig:98038} and \ref{fig:98087}). This improved agreement
with the jet data leads to a slightly
worse description of other distributions such as
fragmentation functions and transverse energy flows (see Figs.
\ref{fig:jane} and \ref{fig:hiq2e}).
Conversely, the new treatment at high $Q^2$ of ARIADNE describes much
better
than before transverse energy flows and event shape variables, but
lessens the agreement with the data on jet shapes. The current
study seems to suggest that a simultaneous description of jet and charged
particle distributions is difficult.

\subsubsection{Approach 2}

The 2nd approach has additionally investigated the behaviour of ARIADNE
for $ Q^2 < 80\ {\rm GeV^2}$ and has also concentrated on
different data sets than those used in 
approach 1. In particular new preliminary data
from H1 on dijet production~\cite{H1dijet} 
has been used along with charged particle
distributions in $\gamma^*P$ centre of mass frame~\cite{H1gamp}.
In addition, the $E_T$
flows and the charged particle distributions in the Breit frame have
been considered but, again, at lower $Q^2$ values than approach 1.
ARIADNE 4.10, with the high $Q^2$ modifications, has been studied using
CTEQ4L for the parton density parametrisation.


Investigations showed that parameter PARA(10) was very
sensitive to the dijet cross section, especially at low $Q^2,$ and was
also sensitive, to a lesser degree, to the $E_T$ flows. Parameter
PARA(25) also influenced the agreement with the dijet measurement and 
to the rapidity distribution of hard $p_T$ charged 
particles but otherwise displayed little sensitivity to the data.
The other two parameters, PARA(15) and PARA(27),
displayed a lesser sensitivity to the data, though the hard $p_T$
particles and the dijet distributions proved the most affected to
changes in these parameters.

\begin{figure}[htb] 
  \vspace*{2mm}
 \begin{center}
 \epsfig{figure=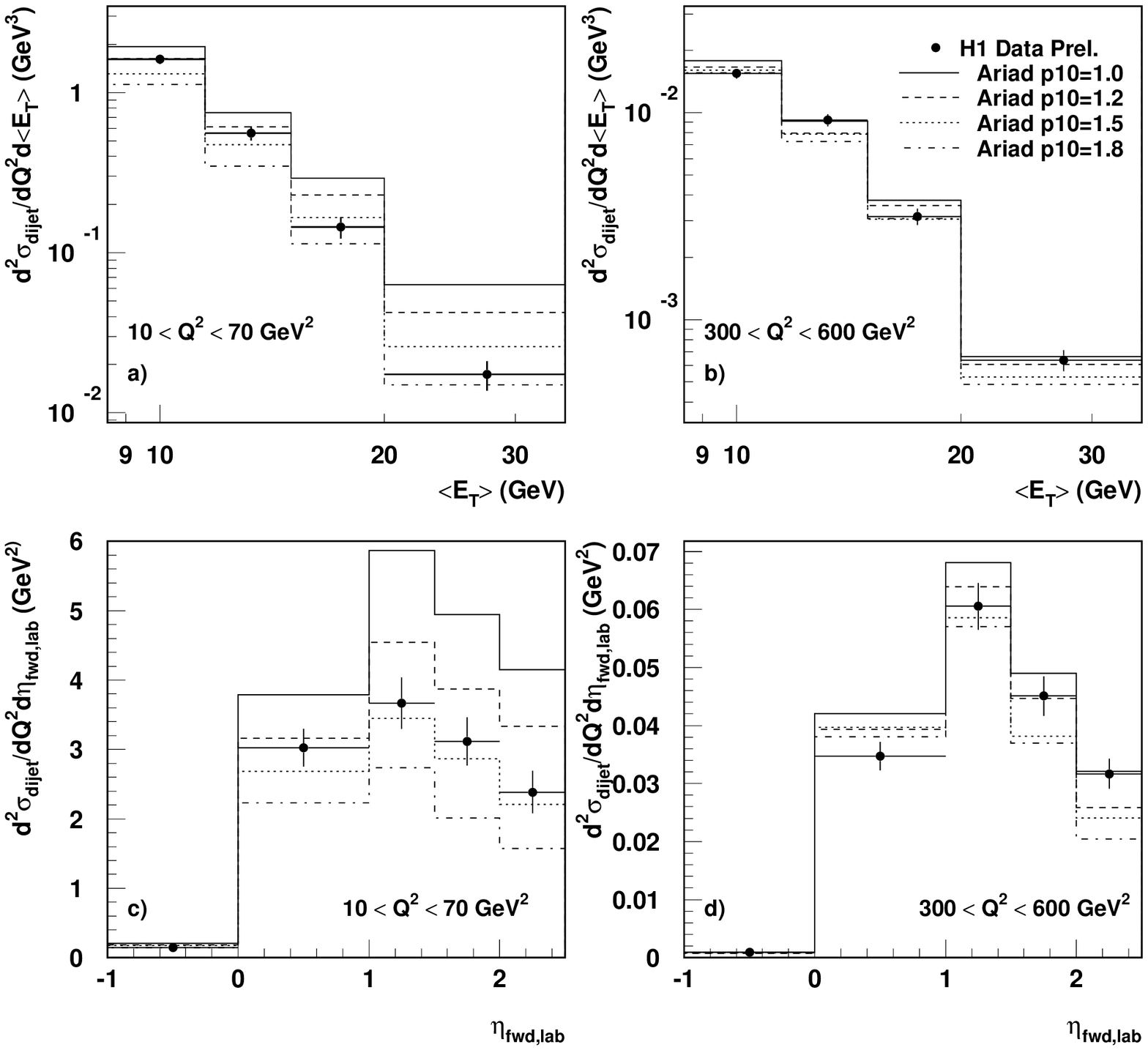,width=15.cm}
 \end{center}
\caption{{\it
Dijet cross section ($E_{T1} + E_{T2} > 17 $ \GeV,
$-1 < \eta_{lab} < 2.5 $) as function of the mean $E_T$
of the jets in the Breit frame and of the pseudo-rapidity
of the forward jet in two bins of $Q^2$.
Shown are preliminary data from the H1 collaboration
and the ARIADNE prediction for various PARA(10) values.
  }}
\label{fig:carli1}
\end{figure}

\begin{figure}[htb] 
  \vspace*{2mm}
 \begin{center}
 \epsfig{figure=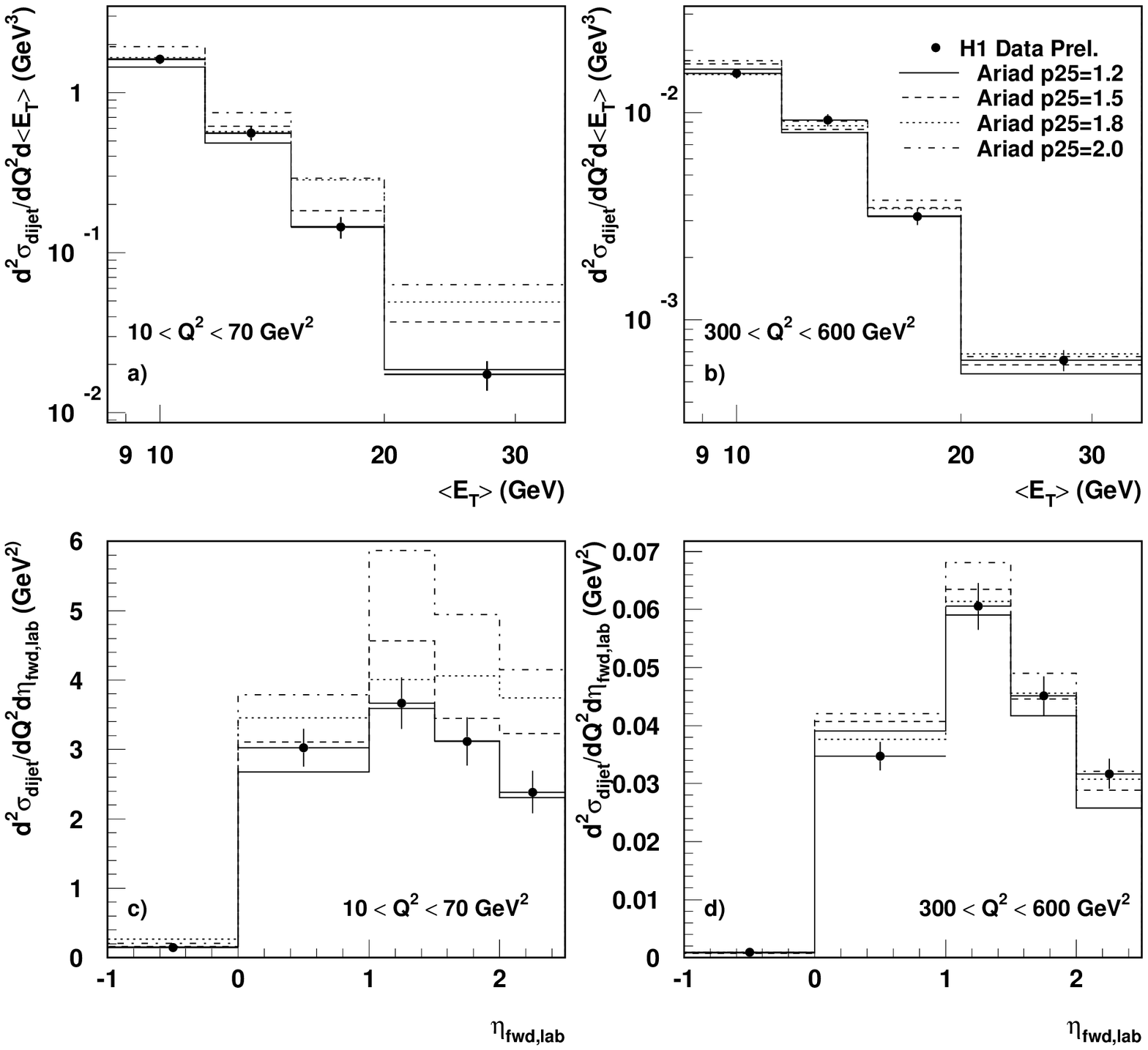,width=15.cm}
 \end{center}
\caption{{\it
Dijet cross section ($E_{T1} + E_{T2} > 17 $ \GeV,
$-1 < \eta_{lab} < 2.5 $) as function of the mean $E_T$
of the jets in the Breit frame and of the pseudo-rapidity
of the forward jet in two bins of $Q^2$.
Shown are preliminary data from the H1 collaboration
and the ARIADNE prediction for various PARA(25) values.
  }}
\label{fig:carli2}
\end{figure}

Figure~\ref{fig:carli1} shows the sensitivity of the dijet cross section
to PARA(10). The default ARIADNE produces $E_T$ spectra for the dijets
that are 
too hard, with the discrepancy predominantly
occurring in the forward region, $\eta_{fwd,lab} > 1.0.$ 
At low $Q^2$ there is a large variation in the $E_T$ spectrum
but PARA(10) affects the distribution at both low and high
$E_T$, which results in this parameter alone
not being able to describe the complete $E_T$ spectra.
This problem can be circumvented by varying PARA(25) in conjunction
with PARA(10). Variation in PARA(25) alone
gives larger fractional changes in the cross section at
large $E_T$ than at smaller values of $E_T$, see Fig.~\ref{fig:carli2}.

\begin{figure}[htb] 
  \vspace*{2mm}
 \begin{center}
 \epsfig{figure=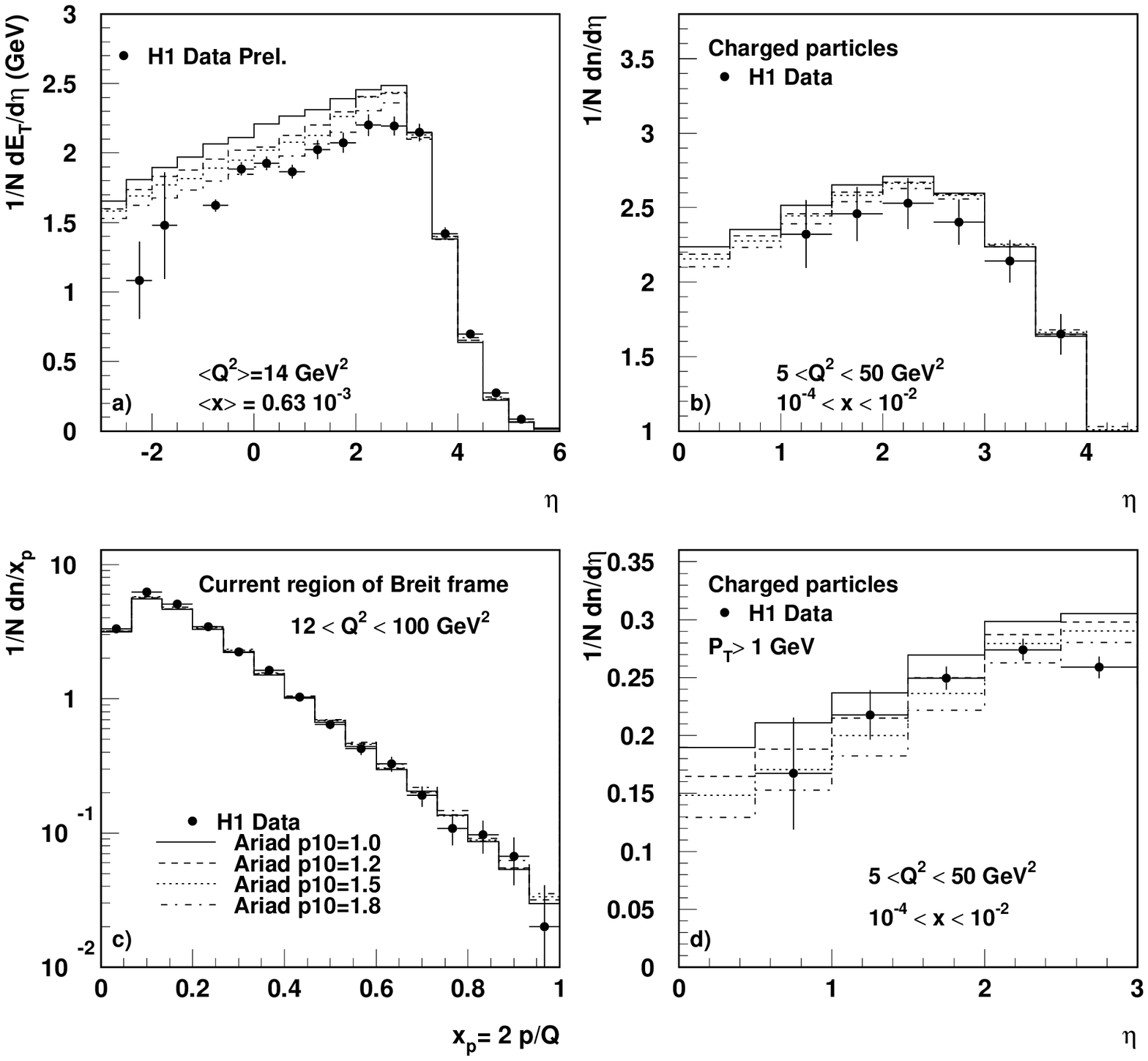,width=15.cm}
 \end{center}
\caption{{\it
The transverse energy as a function of the pseudo-rapidity
$\eta$ (a),
the charged particle multiplicity as a function of $\eta$ (b),
the scaled momentum, $x_p$,
of charged particles in the current region of the
Breit frame (c), and
the multiplicity of hard charged particles
as a function of $\eta$ (d).
Shown are H1 data and the prediction of ARIADNE for
variations of PARA(10).
  }}
\label{fig:carli3}
\end{figure}

\begin{figure}[htb] 
  \vspace*{2mm}
 \begin{center}
 \epsfig{figure=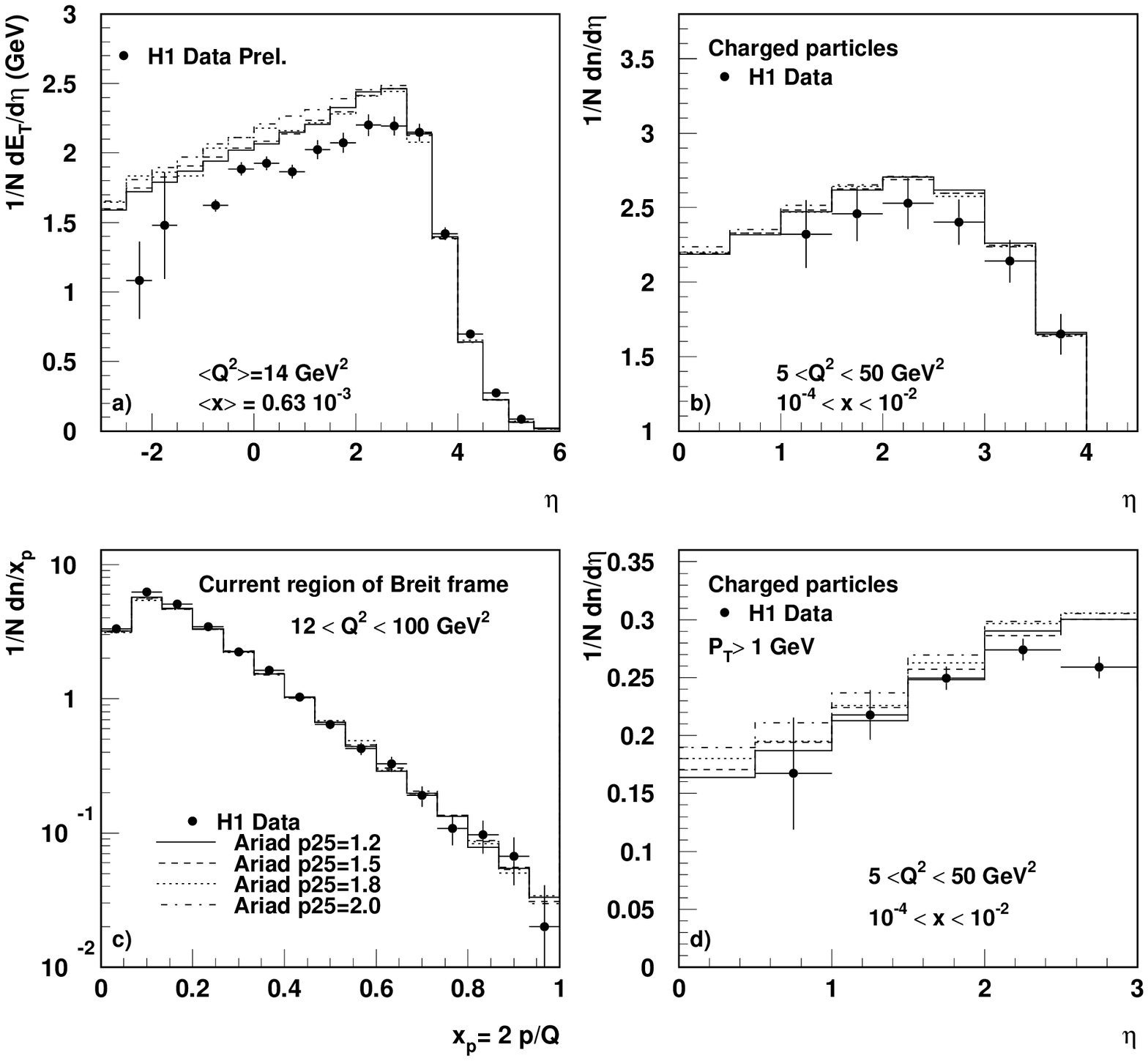,width=15.cm}
 \end{center}
\caption{{\it
The transverse energy as a function of the pseudo-rapidity
$\eta$ (a),
the charged particle multiplicity as a function of $\eta$ (b),
the scaled momentum, $x_p$,
of charged particles in the current region of the
Breit frame (c), and
the multiplicity of hard charged particles
as a function of $\eta$ (d).
Shown are H1 data and the prediction of ARIADNE for
variations of PARA(25).
  }}
\label{fig:carli4}
\end{figure}

The influence of PARA(10) on the $E_T$ flows can be seen in
Fig.~\ref{fig:carli3}. The increase of this parameter suppresses $E_T$
production across the whole $\eta$ range. A similar effect is seen in the
charged particle rapidity
 distribution, particularly for particles with $p_T>1{\rm\ GeV}.$ 
As can be seen from the $x_p$ spectra, Fig.~\ref{fig:carli3}, 
the current region of the
Breit frame seems
relatively insensitive to this parameter.
The $E_T$ flows are less sensitive to PARA(25) than PARA(10), see
Fig.~\ref{fig:carli4}. However the data seem to prefer values of
PARA(25) smaller than the default. This preference is also true for the
charged particle rapidity distributions regardless of any
$p_T$ selection, for the default value of PARA(10).

\begin{table}[hbt]
\centering
\begin{tabular}{|c|c||c|c|c|}
\hline
PARA(10) & PARA(25) & \multicolumn{3}{|c|}{Average $\chi^2$} \\
 & & low $Q^2$ & high $Q^2$ & all $Q^2$ \\
\hline \hline
    & 1.2 & 1.8 & 1.2 & 1.5 \\
1.0 & 1.5 & 2.6 & 1.3 & 2.0 \\
    & 1.8 & 2.6 & 1.3 & 2.0 \\
\hline
    & 1.2 & 2.5 & 1.7 & 2.1 \\
1.2 & 1.5 & 1.6 & 1.3 & 1.5 \\
    & 1.8 & 2.3 & 1.1 & 1.7 \\
\hline
    & 1.2 & 4.1 & 2.6 & 3.4 \\
1.5 & 1.5 & 2.6 & 1.9 & 2.3 \\
    & 1.8 & 2.1 & 1.7 & 1.9 \\
\hline
1.8 & 1.5 & 4.5 & 2.7 & 3.7 \\
    & 1.8 & 3.4 & 2.4 & 2.9 \\
\hline
\end{tabular}
\caption[junk]{{\it Summary of the $\chi^2$ values obtained 
during variation of PARA(10) and PARA(25). (PARA(15) and PARA(27) are
fixed at $1.0$ and $0.6$ respectively.)}}
\label{tab:carli1}
\end{table}

The average $\chi^2$ for the low and high $Q^2$ region, as well as the
combined $Q^2$ regions, are
shown in Table~\ref{tab:carli1} at various settings of PARA(10) and
PARA(25). An improved fit to the data was found for all
distributions for the parameters listed as set 2 in table~\ref{tab:parameters}.
It should be noted though that a comparison with the ZEUS jet shapes was
not included in the data sets investigated in this approach.

\subsection{HERWIG}

\Herwig\ overall has fewer tunable
parameters than the Lund family of
generators \cite{ariadne,lepto,jetset}. In particular the cluster
fragmentation model has far fewer tunable parameters than the Lund 
string model.
Many of the parameters are well constrained by e$^+$e$^-$ annihilation data.
Consequently, those involved with the hard subprocess and the
perturbative QCD evolution of the final state parton shower
were not varied for this study.
It was found previously~\cite{futurephys}, that of the remaining
parameters only a small number 
were seen to have any sensitivity to the distributions
under study in DIS.
Therefore it was decided to
limit this study primarily to the effects of the CLMAX and PSPLT parameters,
where CLMAX relates to the maximum allowed cluster mass and PSPLT is the
exponent in generating the mass distribution of split clusters.

\subsubsection{Approach 1}

The data studied in this approach corresponds to the same data sets
considered in Approach 1 for ARIADNE but also extended to lower $Q^2.$
In addition to the parameters CLMAX and PSPLT, this study investigated
the parameter DECWT, which provides the relative weight
between
decuplet and octet baryon production relevant to the new decay tables.
The dependence on the parton density parameterisation (pdf) of the proton has
also been investigated
by studying CTEQ4L
\cite{cteq4}
and MRSD- \cite{mrsd} pdfs.  Even though the MRSD- has in
principle been
retracted by the authors and is known to be too high at low $x$ it was
used here to
provide a more significant variation of the underlying distribution.

The three parameters were studied over the following ranges:
\begin{eqnarray*}
2.0 < &{\rm CLMAX}& < 5.0 \\
0.6 < &{\rm PSPLT}& < 0.9 \\
0.6 < &{\rm DECWT}& < 0.8.
\end{eqnarray*}

The effect of increasing CLMAX is to increase the $E_T$ flow as does
increasing PSPLT.
Increasing the $E_T$ flow with CLMAX has the effect of broadening the
jet shapes  and producing harder momenta spectra for the
charged particles. This is thought to be due to the fact
that the clusters are allowed
to have more energy before they are forced to split.
Reducing DECWT increases the $E_T$ flow predictions 
at low values of PSPLT with a smaller reduction or slight
increase for larger values of PSPLT.

\begin{figure}[tph]
\centerline{\epsfig{file=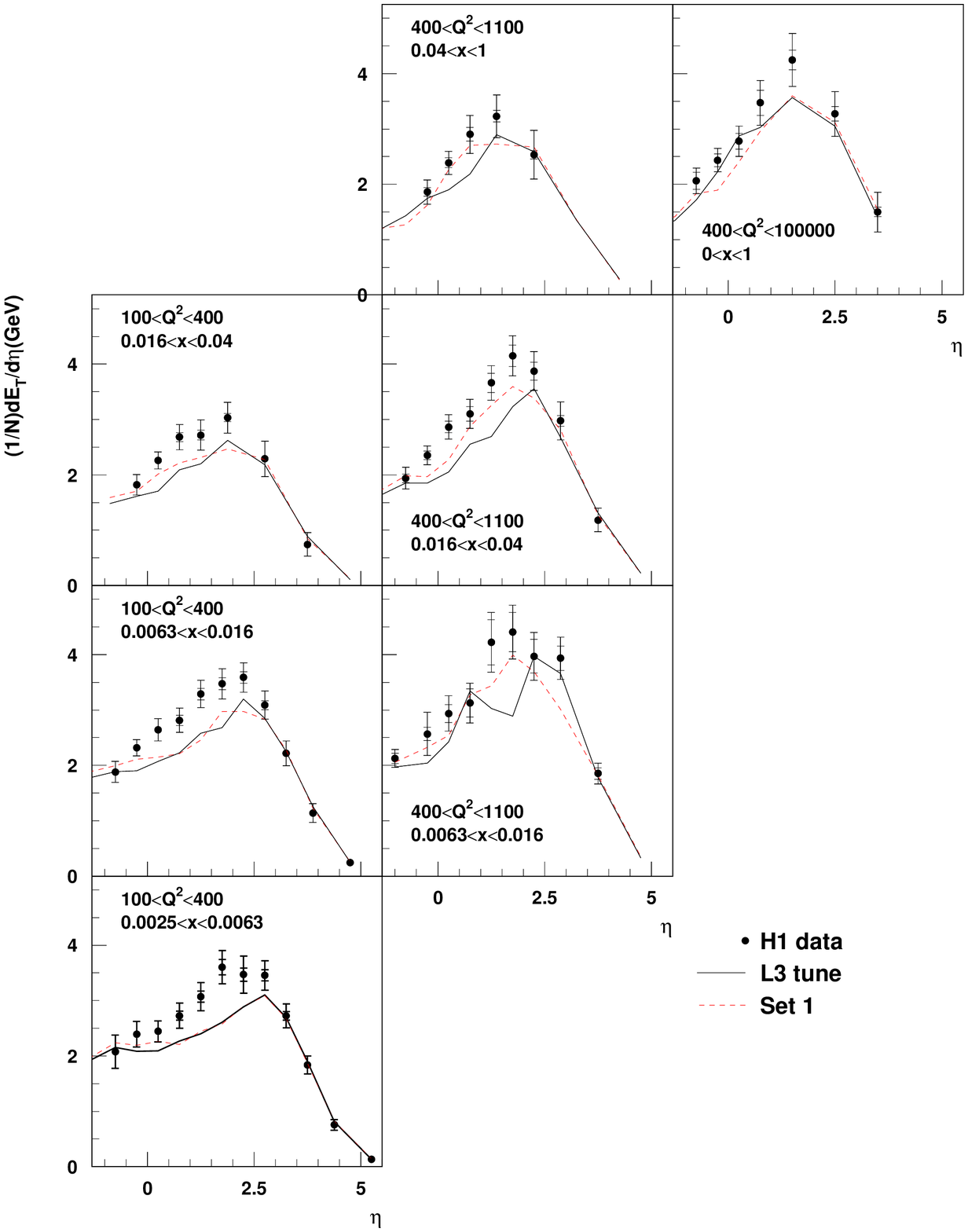,width=0.9\textwidth}}
\caption{{\it H1 preliminary
transverse energy flow data compared to predictions of the
HERWIG generator at various parameter settings (approach 1).}}
\label{HRW1et}
\end{figure}

An attempt was made to tune the standard HERWIG (using MRSD- parton
density functions) and compare with tuned
values from LEP data from L3. The best values of the parameters
achieved for the HERA data are listed in Table~\ref{tab:hrw1}.
Neither the `tuned' set 1 parameters or the L3 parameters can describe the
transverse energy flows at low $x$ and $Q^2$, see Fig.~\ref{HRW1et},
whilst at higher $Q^2$ and $x$ the `tuned' values give a better
description of the HERA data than the
L3 values. The jet shape distributions also prefer the `tuned' values.
With the parameters chosen for investigation it was not possible to
achieve a consistent description of the data at both low and high 
$(x,Q^2).$

\begin{table}[h]
\begin{center}
\begin{tabular}{|l|c|c|c|c||c||c|}
\hline
 & \multicolumn{4}{c||}{HERA} & LEP & \\
Parameter &  set 1 (NLO) & set 2 (LO) & set 3 (LO) & set 4 (LO) &L3 &
default\\
\hline
CLMAX&  4.0  & 5.0 & 3.0 & 4.0 & 3.1 & 3.35 \\
PSPLT&  0.8 & 0.8 & 0.6 & 0.9 & 1.0 & 1.0 \\
DECWT&  0.75 & 0.75 & 0.65 & 0.75 & 0.5 & 1.0 \\
\hline
\end{tabular}
\caption[junk]{{\it Summary of the investigated parameter settings for DIS
HERWIG, with NLO and LO running of $\alpha_s.$ Also shown are the
default parameters and a parameter set find by the L3 collaboration
tuning to LEP data.}}
\label{tab:hrw1}
\end{center}
\end{table}

\begin{figure}[tph]
\centerline{\epsfig{file=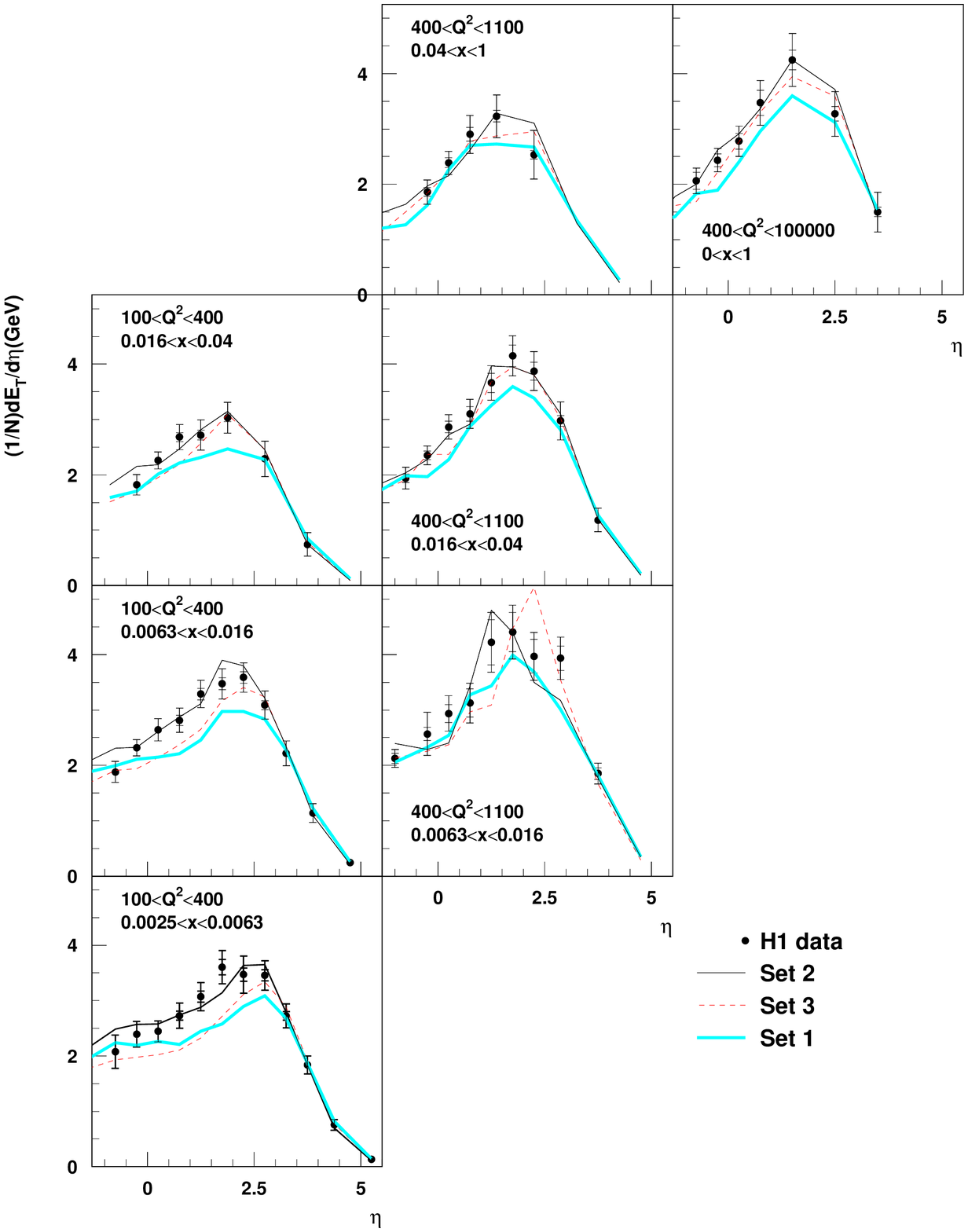,width=0.9\textwidth}}
\caption{{\it H1 transverse energy flow data compared to predictions of the
HERWIG generator at various parameter settings (approach 1).}}
\label{HRW23et}
\end{figure}

\begin{figure}[tph]
\centerline{\epsfig{file=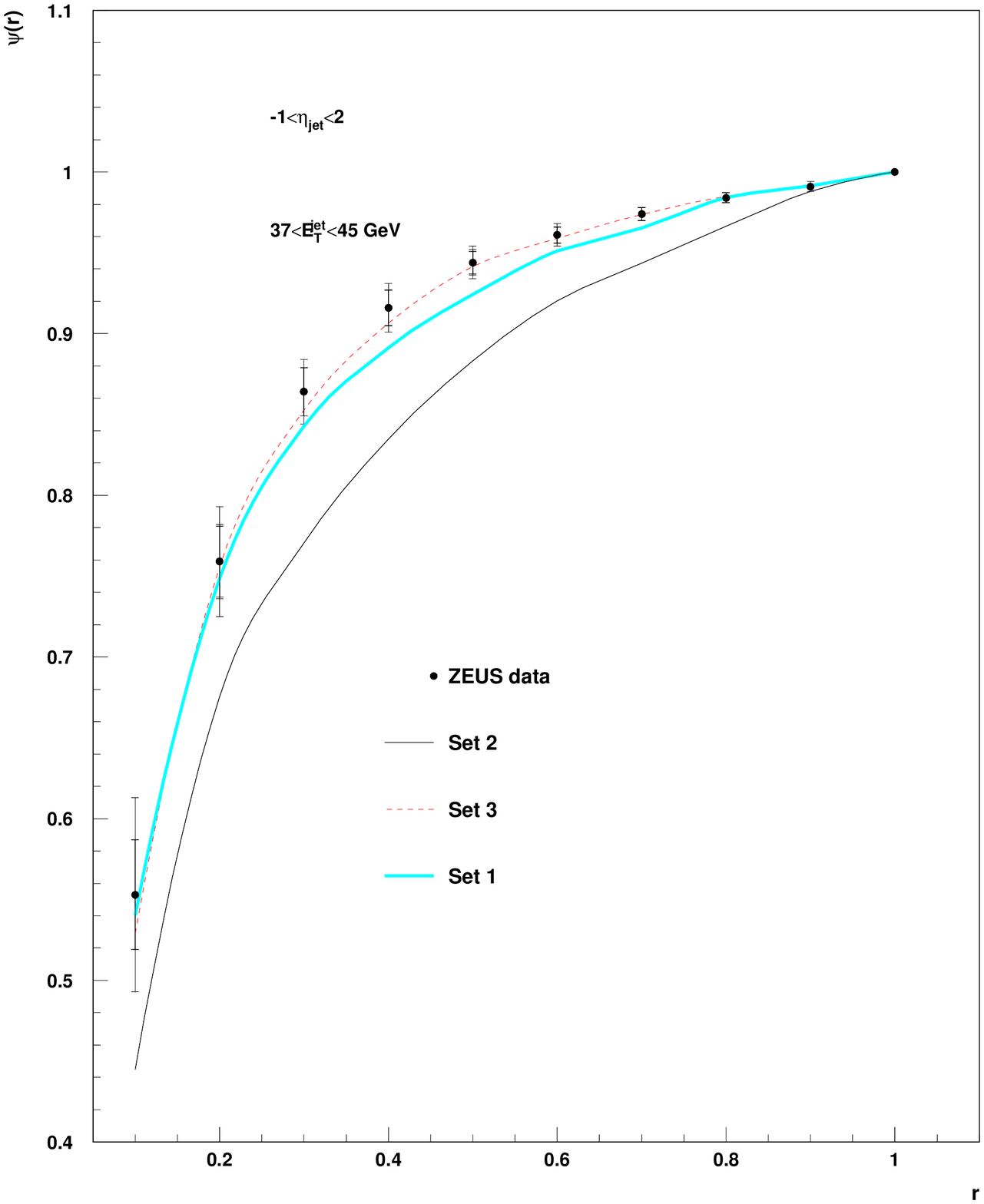,height=0.45\textheight}}
\centerline{\epsfig{file=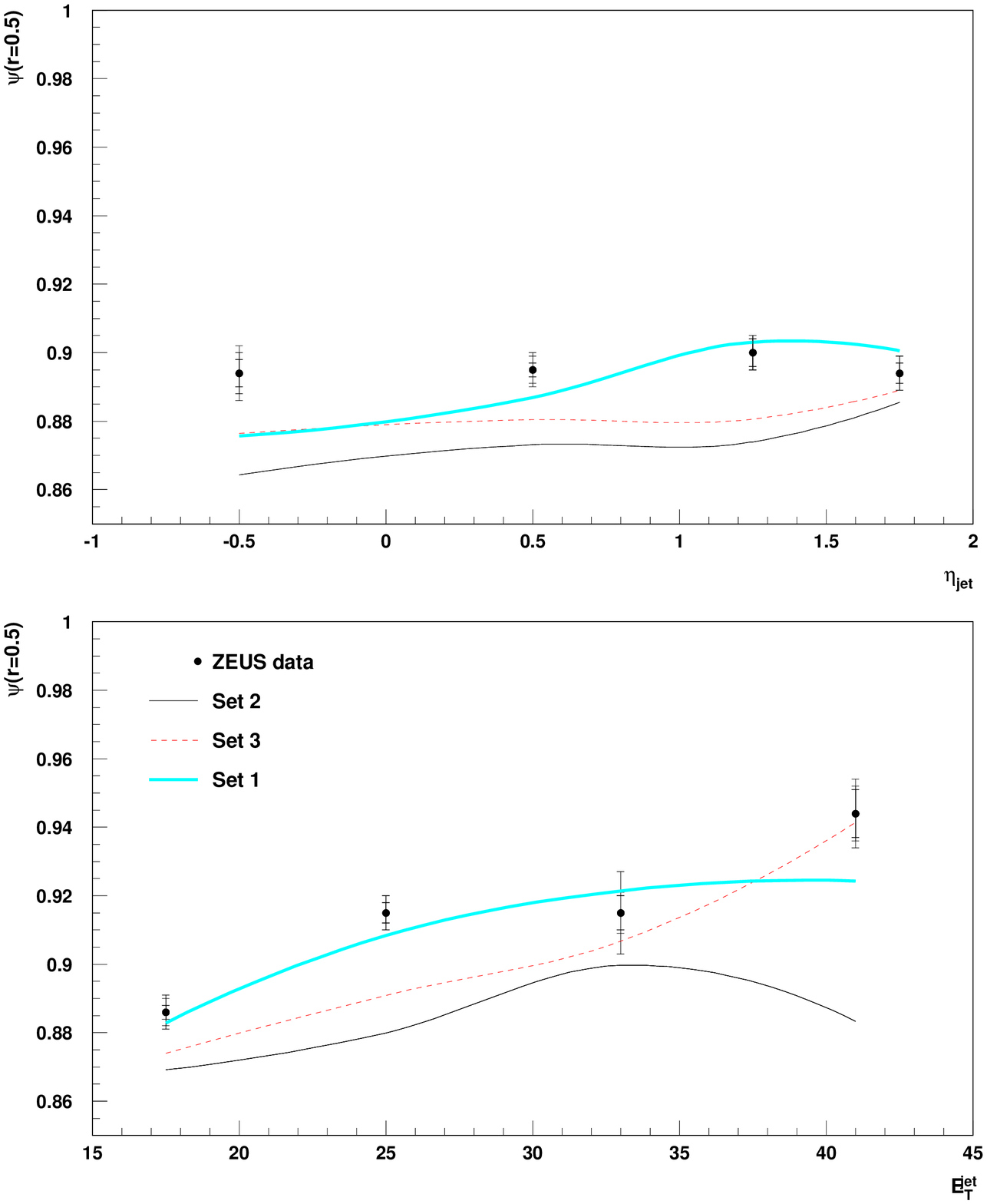,height=0.45\textheight}}
\caption{{\it ZEUS  jet profile data compared to predictions of the
HERWIG generator at various parameter settings (approach 1).}}
\label{HRW23js}
\end{figure}

In an attempt to overcome the difficulty in obtaining sufficient 
$E_T$ at higher $Q^2$ without using very high values of CLMAX, an
investigation of HERWIG with LO running $\alpha_s$ was made.
Two sets of parameter settings are shown in Table~\ref{tab:hrw1} for
this modified HERWIG, in conjunction with MRSD-. 
Set 2 gives the best description of the $E_T$ flows,
whilst conversely set 3 gives a better description of the jet shape data.
Figure~\ref{HRW23et} compares the HERWIG model predictions for the $E_T$
flows with the data. Set 2 describes these distributions well over
the whole $x$ and $Q^2$ range. Set 3 also improves the description
of this data in the highest $Q^2$ bins, though it underestimates the
data in the lowest bins. 
Figure~\ref{HRW23js} compares the HERWIG predictions, with the
parameter sets, to the jet shape data. 
Set 3 gives a better description of
this data than using the set 2 parameters. Set 2 predicts jets
broader than that observed in the data and is in poor agreement with the
data.

\begin{table}[hbt]
\centering
\begin{tabular} {|l|l|c|c|}
\hline
Parameter set & parton density set & \multicolumn{2}{|c|}{Average $\chi^2$}\\
& & $E_T$ flows & $x_p$ \\
\hline
LEP(L3) & MRSD- & 2.78 & 3.56 \\
set 1 & & 1.84 & 3.46 \\
set 2 & & 0.63 & 2.67 \\
set 3 & & 1.41 & 3.40  \\
set 4 & & 0.59 & 2.58 \\
\hline
set 2 & CTEQ4L & 0.76 & 2.79 \\
set 3 & & 1.82 & 3.26 \\
set 4 & & 0.80 & 2.37 \\
\hline
\end{tabular}
\caption[junk]{{\it Summary of the $\chi^2$ values obtained before and
after tuning for all sets of hadronic final state data.}}
\label{tab:sooty}
\end{table}

Investigation of the sensitivity of the data (and the subsequent
parameter settings) to the choice of parton density functions was made
in the modified HERWIG for the MRSD- and CTEQ4L parametrisations.
In particular the $E_T$ flows and the jet shapes were sensitive 
to the choice of
parton densities. A 4th parameter set was found using the
CTEQ4L parton densities. Again a consistent description of both the $E_T$
flows and the jet shapes was not possible. The parameter set listed in
table~\ref{tab:sooty}
gave a better description of the $E_T$ flows than the jet shapes.
The $\chi^2$ achieved for the various parameter settings of HERWIG using
both MRSD- and CTEQ4L are
given in Table~\ref{tab:sooty} for the $x_p$ distribution in the current
region and the $E_T$ flows.

\subsubsection{Approach 2}

This study considered the same data samples as approach 2 for ARIADNE.
Only the H1 modified HERWIG, with the running of $\alpha_s$ at leading
order, has been considered. The parton densities used in this approach
correspond to CTEQ4L.

\begin{figure}[htb] 
  \vspace*{2mm}
 \begin{center}
 \epsfig{figure=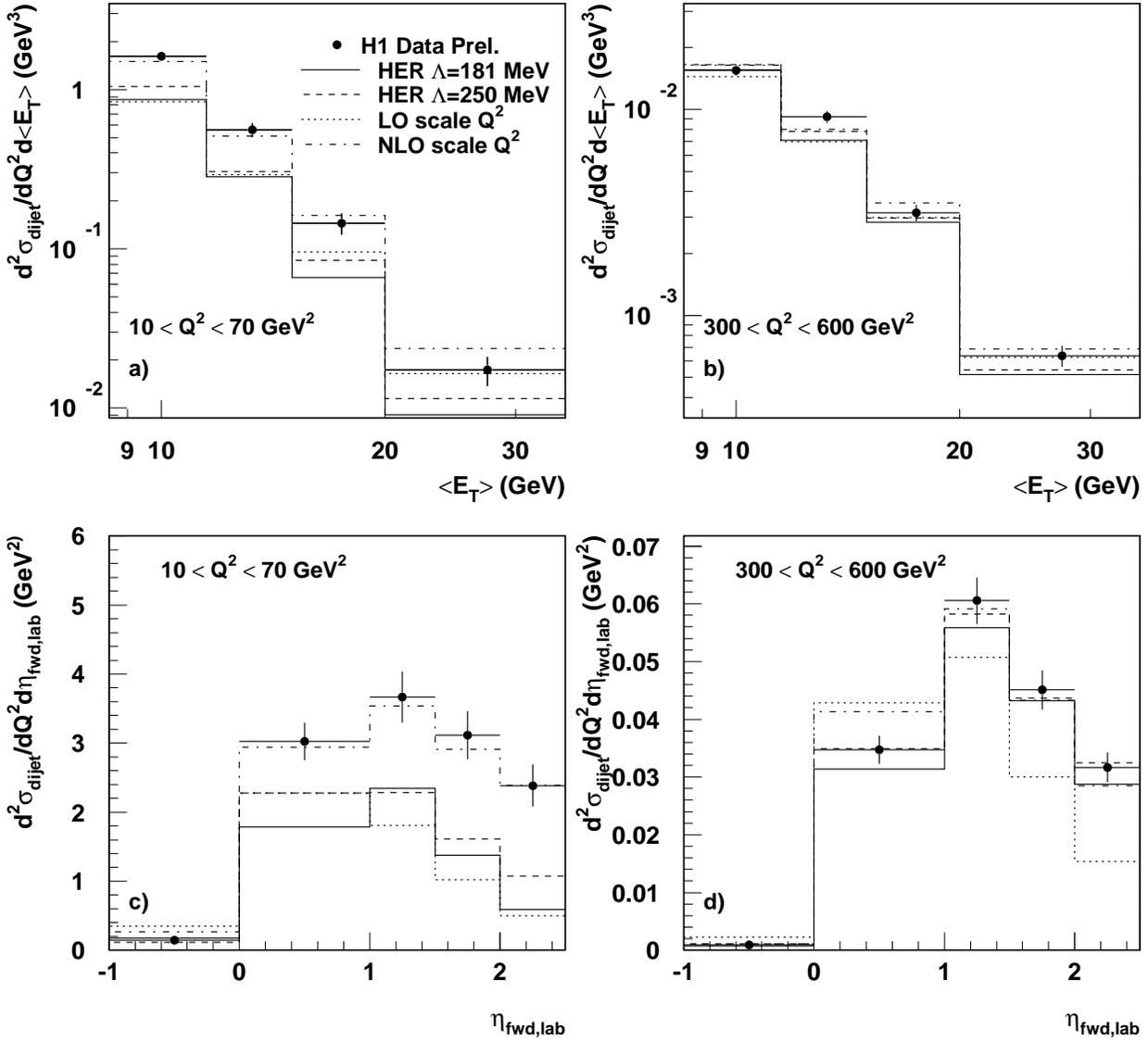,width=\textwidth}
 \end{center}
\caption[Herwig Jets]{{\it
Dijet cross section ($E_{T1} + E_{T2} > 17 $ \GeV,
$-1 < \eta_{lab} < 2.5 $) as function of the mean $E_T$
of the jets in the Breit frame and of the pseudo-rapidity
of the forward jet in two bins of $Q^2$.
Shown are preliminary data from the H1 collaboration
and the prediction of HERWIG and the DISENT.
The DISENT prediction is given in ${\cal O}(\alpha_s)$ (LO)
${\cal O}(\alpha_s^2)$ (NLO) for $Q^2$ as renormalisation
scale.  }}
\label{fig:carli_hrw1}
\end{figure}

At high $Q^2$ a reasonable description of the dijet data by HERWIG could
 be obtained only if a larger (than default) value of
$\alpha_s$ ($\Lambda = 250 {\rm MeV}$) was used, Figure~\ref{fig:carli_hrw1}. 
At low $Q^2$ HERWIG was unable to achieve a good
description of the dijet data. The dijet cross sections were relatively
insensitive to changes in the hadronisation parameters.

In Figure~\ref{fig:carli_hrw1} the 
DISENT\footnote{The DISENT
program incorporates a NLO calculation for DIS at the parton level. It
can also be used to obtain partonic LO predictions.}~\cite{disent}
 ${\cal O}(\alpha_s)$ predictions
(using $Q^2$ as the renormalisation scale) are compared to HERWIG.
The HERWIG predictions are in agreement with the 
DISENT LO calculation.
In the same figure,
it is also shown that the NLO  corrections (K-factors), in
particular at low $Q^2$ and forward pseudorapidities 
$\eta_{\rm fwd,lab} \sim 2,$ are large.
The parton showers used to emulate higher
orders in HERWIG are insufficient to account for these large NLO
corrections.

\begin{table}[hbt]
\centering
\begin{tabular}{|c|c||c|c|c|}
\hline
PSPLT & CLMAX & \multicolumn{3}{|c|}{Average $\chi^2$} \\
 & & low $Q^2$ & high $Q^2$ & all $Q^2$ \\
\hline \hline
0.5 & 3.0 & 14.5 & 3.0 & 9.5 \\
    & 5.0 & 10.0 & 5.0 & 7.7 \\
\hline
0.65& 3.0 & 14.1 & 2.6 & 8.8 \\
    & 5.0 &  8.8 & 4.1 & 6.7 \\
\hline
0.8 & 3.0 & 15.3 & 2.5 & 9.5 \\
    & 5.0 &  7.9 & 4.0 & 6.1 \\
\hline
1.0 & 3.0 & 18.3 & 1.9 & 10.8 \\
    & 5.0 &  7.5 & 3.4 & 5.6 \\
\hline
1.2 & 3.0 & 23.0 & 1.7 & 13.3 \\
    & 5.0 & 10.5 & 3.3 & 7.2 \\
\hline
\end{tabular}
\caption[junk]{{\it Summary of the $\chi^2$ values obtained 
during variation of PSPLT and CLMAX.}}
\label{tab:carli_hrw1}
\end{table}

\begin{figure}[htb] 
  \vspace*{2mm}
 \begin{center}
 \epsfig{figure=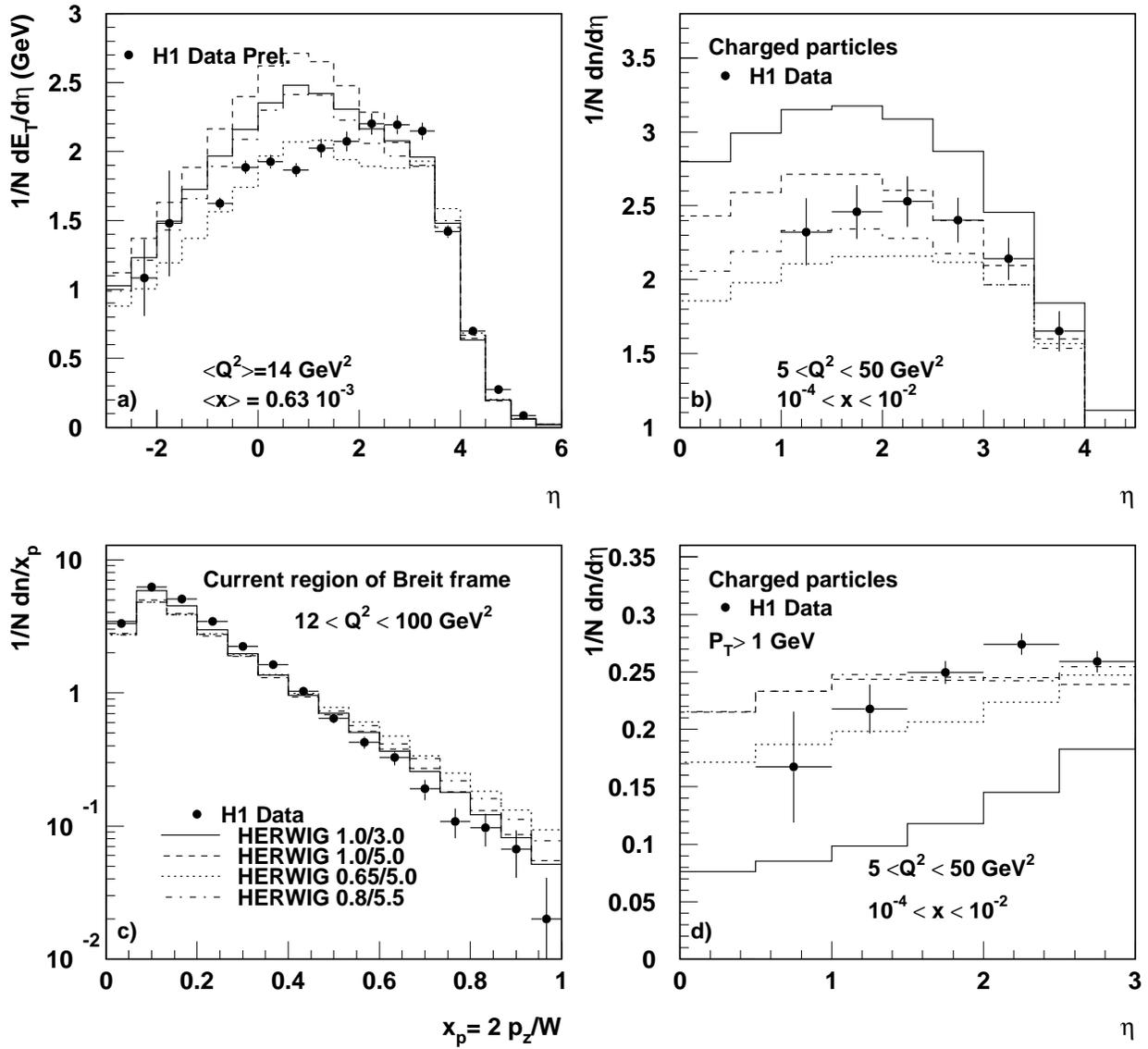,width=\textwidth}
 \end{center}
\caption[Herwig mult]{{\it
The transverse energy as a function of the pseudo-rapidity
$\eta$ (a),
the charged particle multiplicity as a function of $\eta$ (b),
the scaled momentum, $x_p$,
of charged particles in the current region of the
Breit frame (c), and
the multiplicity of hard charged particles
as a function of $\eta$ (d).
Shown are H1 data and the prediction of HERWIG for
various settings of fragmentation parameters
PSPLT and CLMAX.  }}
\label{fig:carli_hrw2}
\end{figure}

In contrast to the dijet data,
the $E_T$ flows and the rapidity distributions of charged particles
exhibit a strong dependence on the fragmentation parameters. 
As in approach 1, the current region of the Breit frame and the high $Q^2$ data
prefer different settings of PSPLT and CLMAX parameters than does the low
$Q^2$ data. The results are summarised in
Table~\ref{tab:carli_hrw1} and the HERWIG predictions are compared
to the data in figure~\ref{fig:carli_hrw2}.
The high $Q^2$ and the Breit frame current region
data prefers settings of ${\rm CLMAX}=3.0$ (the default)
and ${\rm PSPLT}=1.2$
whilst the low $Q^2$ data favour a higher value of ${\rm CLMAX}=5.0$
with a slightly lower value of ${\rm PSPLT}=1.0.$

Although variation of the fragmentation parameters leads to large
changes in the prediction of the HERWIG model, the underlying parton
dynamics in HERWIG are not sufficient to describe the HERA DIS data.

\subsection{LEPTO}
The new version (6.5.2$\beta$) of LEPTO was confronted with preliminary high
statistics $(2+1)$
jet data from the H1 collaboration~\cite{H1newjets} (statistical error
only on the data.) 
This data set consists of DIS events
which are all forced to be of a
$(2+1)$ jet configuration  using the modified
Durham algorithm. The distributions studied were $y_2,$ the cut-off
in the algorithm 
where an event is first
defined as $(2+1)$ and the angles in the laboratory frame of the
forward and backward going jet ($\theta_{\rm fwd}$ and $\theta_{\rm bwd}$). 
In addition the jet variables $x_{\rm jet}$, defined as $Q^2/(Q^2+\hat s)$ where
$\hat s$ is the invariant mass of the jet(parton) pair, and $z_p$, defined as
$1/2(1-\cos\theta^\ast)$ where $\theta^\ast$ is the polar angle of jet
in the photon-parton centre of mass system, were investigated.

The following parameters, that control the cut off in the
${\cal O}(\alpha_s)$ matrix element in the generator,
were found to have significant impact on the description of the
data and
 have been studied:
\begin{itemize}
\item PARL(8) $z_p^{min}$ cut off, and
\item PARL(9) $\hat s^{min}$ cut off.
\end{itemize}

\begin{table}[h]
\centering
\begin{tabular}{|c||c|c|c|c|}  \hline
Var. & {\footnotesize LEPTO 6.5} &  {\footnotesize LEPTO
6.5.2$\beta$} &
{\footnotesize LEPTO 6.5.2$\beta$}&  {\footnotesize LEPTO 6.5.2$\beta$}\\
         & default   & default   & set 1 & set 2
\\ \hline\hline

$y_2$  & 517.30& 89.40& 9.54&11.57\\
$\theta_{\rm fwd}$  & 630.04& 97.41& 36.52& 31.43\\
$\theta_{\rm fwd}$  & 499.88& 99.87& 16.91& 14.45\\
$z_p$  & 765.81& 134.74& 82.22& 79.89\\
$x_{\rm jet}$  & 752.58& 175.35& 66.65& 82.10\\
\hline
\end{tabular}
\caption{{\it Summary of the $\chi^2/d.o.f.$ for LEPTO. }}
\label{tab:marc1}
\end{table}

\begin{figure}[tph]
\centerline{\epsfig{file=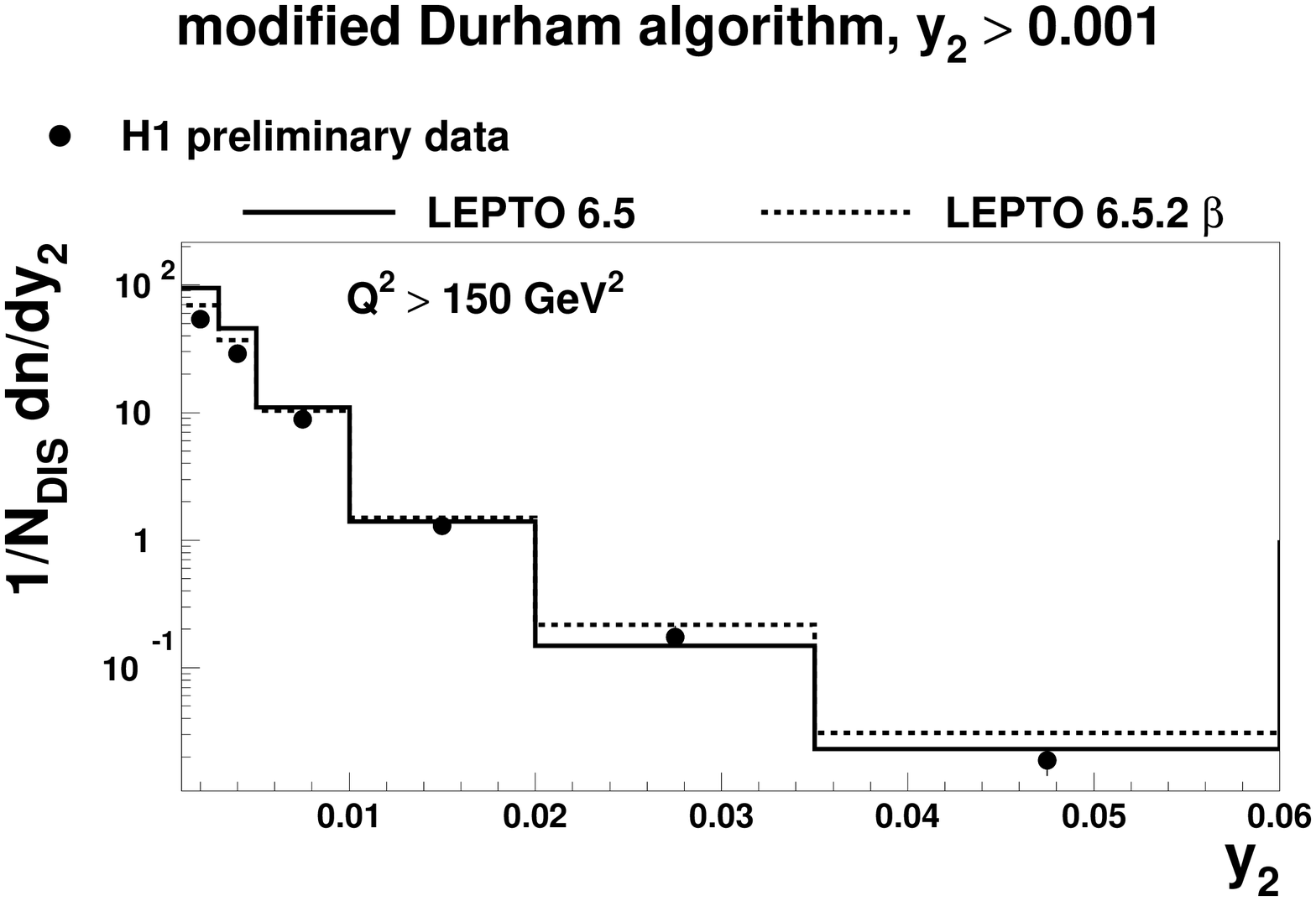,width=0.9\textwidth,bbllx=0pt,bblly=0pt,bburx=650pt,bbury=430pt}}
\caption{{\it H1 preliminary jet data for the variable $y_2$ compared to
pre-workshop version of LEPTO (6.5) and the version developed with SCI
suppression (6.5.2$\beta$)}}
\label{fig:marc1}
\end{figure}

\begin{figure}[tph]
\centerline{\epsfig{file=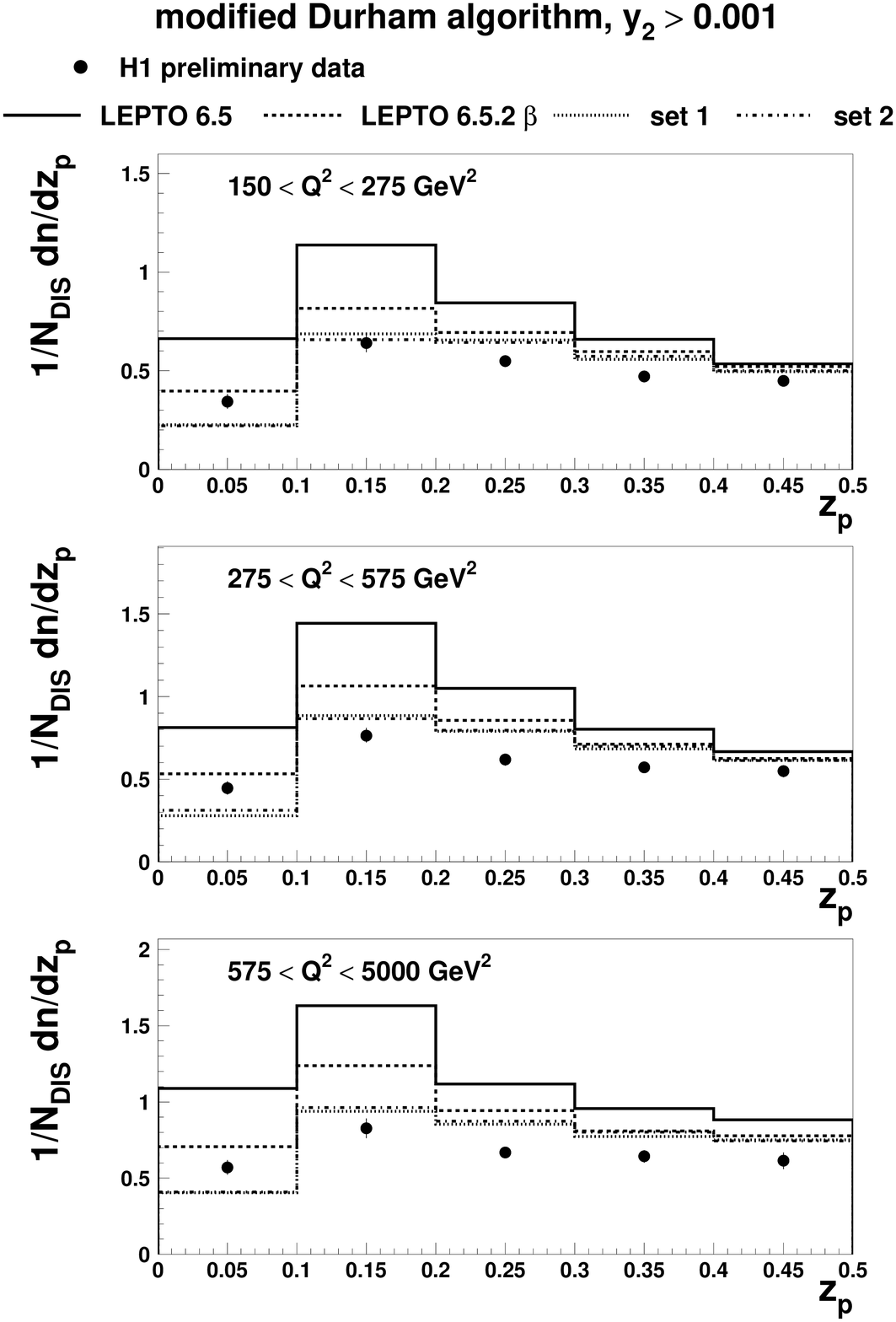,width=0.9\textwidth,bbllx=0pt,bblly=0pt,bburx=665pt,bbury=990pt}}
\caption{{\it H1 preliminary jet data for the variable $z_p$ compared to
pre-workshop version of LEPTO (6.5), the default version developed with SCI
suppression (6.5.2$\beta$) and 2 parameter sets for the new version derived from
this preliminary jet data.}}
\label{fig:marc2}
\end{figure}

The new SCI scheme implemented
in LEPTO version 6.5.2$\beta$ leads to a dramatic improvement in the description
of the data compared with version 6.5. The $\chi^2$ is typically reduced
by a factor of 5--6, see Table~\ref{tab:marc1}. The predictions of LEPTO
compared to the data are shown
in Figures~\ref{fig:marc1} and~\ref{fig:marc2}. 
Further significant improvement in the description of the data with LEPTO
6.5.2$\beta$ was achieved by optimizing the parameters PARL(8) and PARL(9).
The results of this optimization are shown
in Table~\ref{tab:marc2} (set 1) and the improvement can clearly be seen
in the comparison with the data in Figure~\ref{fig:marc2}; the
corresponding $\chi^2$ values are given in table~\ref{tab:marc1}.
It should also be noted that LEPTO describes the $Q^2$ dependence of the
jet distribution well.

\begin{table}[h]
\centering
\begin{tabular}{|l|c|c|}  \hline
          & PARL(8) & PARL(9) \\ \hline\hline

LEPTO 6.5:  default  & 0.04 & 25.0 \\ \hline
LEPTO 6.5.2$\beta$: default &  0.04 & 25.0 \\ \hline
LEPTO 6.5.2$\beta$: set 1 &  0.10 & 25.0 \\ \hline
LEPTO 6.5.2$\beta$: set 2 &  0.04 & 1.0  \\ \hline

\end{tabular}
\caption{{\it LEPTO parameter sets}}
\label{tab:marc2}
\end{table}

A complementary way to optimize LEPTO for jet distributions, instead
of applying hard cuts on divergences of the matrix element (set 1), is
to loosen these cuts so that LEPTO is forced to find appropriate
divergency cuts on an event--by--event basis. 
The preferred values of PARL(8) and PARL(9) using this approach are
listed in Table~\ref{tab:marc2} (set 2) and the corresponding $\chi^2$
values in Table~\ref{tab:marc1}.

The variation of the intrinsic $k_T$, PARL(3), and the cut--off
value of the initial--state parton shower, PYPAR(22), 
had no effect on the quality of the description of the jet data. 
Also the jet data were insensitive to the choice of the parton density 
functions.

Although both approaches to describing the data with LEPTO,
via PARL(8) and PARL(9), result in  significant improvements,
no satisfactory description of the measured 2--jet distributions
could be achieved.  The parameter sets 1 and 2 were then cross checked
against the data samples used in approach~1 for ARIADNE but over the whole
$Q^2$ range, see Table~\ref{tab:marc3}. Besides the $(2+1)$ jet rate and
the charged particle $x_p$ distribution in the current fragmentation
region, the default version of LEPTO 6.5.2$\beta$ gave a better description of the
data.

\begin{table}[htbp]
\centering
  \begin{tabular}{|c|c|c|c|}
  \hline
  Data set ref. & $\chi^2_{6.5.2\beta}$ & $\chi^2_{\rm set 1}$ & $\chi^2_{\rm set 2}$\\
  \hline
  \hline
  \cite{jane}  &  13.56 &  16.6 & 9.32 \\
  \hline
  \cite{hiq2e} &  4.91 &  14.1 & 8.93 \\
  \hline
  \cite{97098} &  1.79 &  4.50 & 2.71 \\
  \hline
  \cite{97108} &  2.13 &  3.08 & 1.53 \\
  \hline
  \cite{98087} &  2.41 &  1.64 & 3.14 \\
  \hline
  \end{tabular}
\caption[junk]{{\it Summary of the $\chi^2/d.o.f.$ values obtained before and
after tuning of LEPTO 6.5.2$\beta$ with jet data for 
various sets of hadronic final state data not used in the tuning.}}
\label{tab:marc3}
\end{table}

\section{Summary}

During the course of the workshop new versions of the LEPTO and ARIADNE
Monte Carlo generators were made available. These modified versions of
the generators were in far better agreement with data.

An attempt was made to find  sets of parameters for the ARIADNE, LEPTO
and HERWIG generators that would describe the DIS HERA data. It proved
difficult to find such a parameter set that would describe the whole range of
distributions at both low and high $Q^2.$ A number of parameter sets are
given for each generator that are optimised for a particular region of
phase space.

This paper attempts to summarise a `snapshot' of
an ongoing program of work between experimentalists of both the H1 and the
ZEUS collaborations and the authors of the event generators.  The
ultimate aim is to have event generators that are able to describe the
complex structure of DIS events at HERA as impressively as they do the LEP data.

\newpage

\newpage
\newpage

\newpage

\newpage

\newpage

\newpage

\newpage

\newpage

\newpage

\newpage

\newpage

\newpage

\newpage

\end{document}